# Novel inverse multi-objective optimization-empowered design of microperforated panels for enhanced low-frequency noise mitigation

Duo Zhang, Ph. D.
Postdoctoral Researcher
Email: duo-dz.zhang@polyu.edu.hk
Department of Civil and Environmental Engineering
The Hong Kong Polytechnic University
Hong Kong, China

Yang Zhang
Ph.D. Student
Email: yang.3.zhang@uconn.edu
Department of Mechanical Engineering
University of Connecticut
Storrs, CT 06269, USA

Sichen Yuan, Ph. D.
Assistant Professor
Email: sichen.yuan@ua.edu
Department of Aerospace Engineering and Mechanics
The University of Alabama
Tuscaloosa, AL 35487, USA

Jiong Tang, Ph. D.
Pratt & Whitney Endowed Professor
Email: jiong.tang@uconn.edu
Department of Mechanical Engineering
University of Connecticut
Storrs, CT 06269, USA

Kai Zhou†, Ph.D. (Corresponding author)
Assistant Professor
Email: cee-kai.zhou@polyu.edu.hk
Department of Civil and Environmental Engineering
Research Institute for Sustainable Urban Development (RISUD)
The Hong Kong Polytechnic University
Hong Kong, China

# Novel inverse multi-objective optimization-empowered design of microperforated panels for enhanced low-frequency noise mitigation

Duo Zhang[a], Yang Zhang[c], Sichen Yuan[d], Jiong Tang[c], and Kai Zhou[a,b] †

a: Department of Civil and Environmental Engineering, The Hong Kong Polytechnic University, Hong Kong, China

b: Research Institute for Sustainable Urban Development (RISUD), The Hong Kong Polytechnic University, Hong Kong, China

c: Department of Mechanical Engineering, University of Connecticut, Storrs, CT 06269, USA

d: Department of Aerospace Engineering and Mechanics, The University of Alabama, Tuscaloosa, AL 35487, USA

## Abstract

Microperforated panels (MPPs) display excellent capacity in noise control applications owing to their high strength, simple design, and efficacy in low-frequency sound absorption. Traditionally, the development of MPPs has relied on a trial-and-error design approach. Although simple optimization-based methods have recently begun to be employed, these designs often overlook practical considerations, such as the increased costs associated with adding more MPP layers, which presents a gap to achieve the practical feasibility of MPP deployment. To address this, the study aims to develop an inverse multi-objective optimization-empowered framework for MPP design to enhance low-frequency noise mitigation while minimizing fabrication costs. Specifically, a finite element (FE) model is established to conduct the acoustic analysis of MPPs, followed by thorough experimental validation. A novel multi-objective particle swarm optimization algorithm (MOPSO) is then developed to cope with mixed-type design variables with interrelations inherent to the MPP architecture. Using the high-fidelity FE model as a cornerstone, the MOPSO guides the inverse optimization analysis to yield multiple non-dominant solutions. These solutions not only avoid the trap of local optima, but also allow for continuous screening to ensure the engineering viability based on empirical judgment. The results clearly demonstrate the effectiveness of the proposed methodology. The MPPs designed in this study show great potential for mitigating indoor noise in buildings, addressing noise issues arising from rapid urbanization and transportation development. Furthermore, the novel optimization strategy proposed in this study holds wide applicability for other sound absorption materials.

## 1. Introduction

Acoustic comfort is a critical concern for urban communities. Rapid urbanization and transportation development have led to severe noise pollution in indoor environments [1,2]. Indoor noise pollution can originate from both indoor and outdoor sources and often has low-frequency characteristics. For instance, indoor noise generated and transmitted by passenger vehicles is around 1 kHz, while the frequencies between 500 Hz and 1 kHz are accompanied by heavy trucks [3]. Additionally, indoor noise can be induced by heating, ventilation, and air conditioning (HVAC) systems propagating through ductwork [4]. The dominant octave band frequency of HVAC noise is even lower, below 400 Hz [5,6]. This noise adversely affects people's lives and work [7,8]. Long-term exposure to severe noise can even lead to various physiological and psychological diseases [9-11]. Therefore, it is important to mitigate indoor noise using effective sound absorbers.

Commercial sound absorption materials can be generally classified into porous sound absorption materials (PMs) and resonant sound absorption materials (RMs) [12]. PMs are widely used to reduce noise across a broad frequency range with high absorption coefficients through air flow resistances, and viscous and thermal losses [13]. However, they are less effective at low frequencies due to the long sound wavelengths [14]. RMs mainly include Helmholtz resonators (HRs) [15] and membrane absorbers [16], which exhibit good absorption properties in low and narrow frequency ranges because of the strong internal resonance effect [17]. Membrane absorbers typically consist of mass blocks and a tensioned membrane [18]. However, the instability of mass blocks and membrane creep result in weak robustness, limiting the practical application of membrane resonators [19]. In contrast, HRs are more likely to be adopted in practice. For improved sound absorption performance and easier manufacturing, HRs have evolved into a new type of sound absorbers, known as perforated panels, which operate on the principle of Helmholtz resonance absorption [20]. By adjusting the perforation ratio, perforated panels and HRs are interchangeable [21]. However, ordinary perforated panels with large holes and high perforation ratios must be used with porous materials in the cavity due to their inherently low acoustic resistance. To address this issue, Dah-You Maa [22] proposed reducing the perforation diameter to enhance acoustic resistance, leading to the development of microperforated panels (MPPs). Compared to ordinary perforated panels, MPPs offer a broader sound absorption band due to the viscous effect around the tiny, perforated holes [1,23].

MPPs can be fabricated from any sheet material. Because of their low cost, high strength, simple design, and efficacy in sound absorption [24], MPPs have been extensively applied to enhance the acoustic environment in various scenarios. Examples include deploying MPPs over the front rotor of the contra-rotating fan [25], inside helicopter cabins [26], in wind tunnel test aeras [27], and in fighter cockpits [28]. MPPs must be installed in front of the rigid wall because their sound absorption performance will be negligible without the air gap [20,29]. This strategic placement endows MPPs with substantial potential to address the indoor noise issues [1], allowing for the design adjustment to adapt to different room settings [30,31]. Several successful cases of indoor noise mitigation using MPPs are shown in Fig. 1.

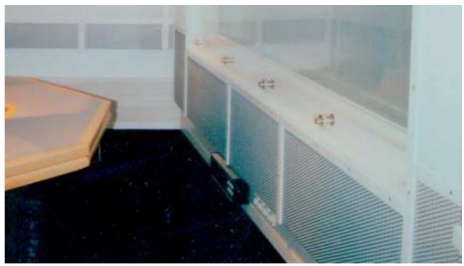

(a)

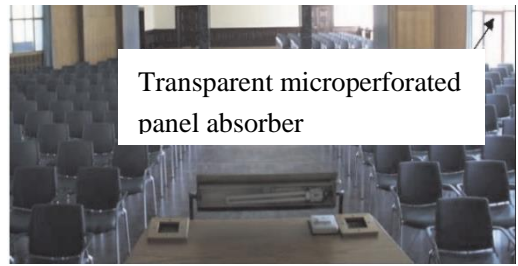

(b)

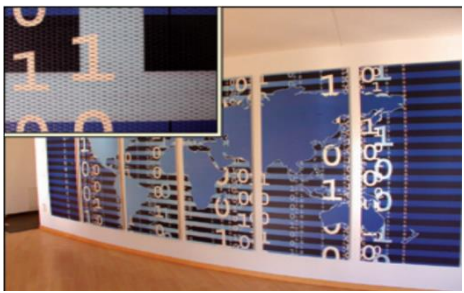

(c)

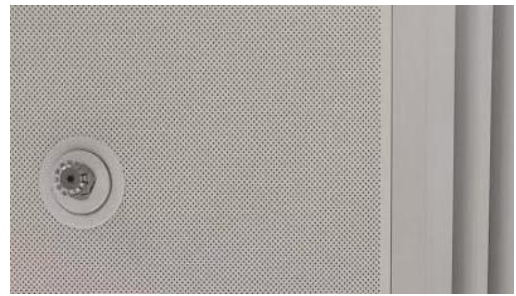

(d)

**Fig.1.** Practical application scenarios of MPPs: (a) a speaker's studio at RTL, Köln, reprinted with permission from [32], copyright (2006) European Acoustics Association; (b) the assembly hall of the University of Freiburg, reprinted with permission from [33], copyright (2001) European

Acoustics Association; (c) MPPs used in the reception area, reprinted with permission from [34], copyright (2011) Tech Science Press; and (d) MPPs around the vent of Pao Yue-kong Library.

Single-layer MPP is commonly used for noise reduction. However, it is limited in providing a broad frequency attenuation band compared to conventional porous materials [35]. Conversely, multi-layer MPPs, comprised of their different air cavities, can extend the frequency attenuation band by enabling multiple absorption peaks. However, adding layers also introduces additional air cavities, which increases the thickness of synthesized MPPs. This may lead to installation challenges in practical applications. Furthermore, the assembly process becomes more complex, and the associated costs can rise significantly. From a design standpoint, the design of a multi-layer MPP requires the examination of many influential structural parameters and their intrinsic intercorrelation, making it technically challenging. Currently, existing studies performed the MPP design based on the trial-and-error approach, which heavily relies on experience and judgment [36-40]. For a single-layer configuration, this approach might be feasible because of the limited number of structural parameters being investigated during design process. However, due to the increasing complexity of the multi-layer MPP design mentioned above, the trial-and-error approach struggles to balance design accuracy with time efficiency [41].

With the advancement of computational power, computational design has emerged as an effective solution to advance MPP development. In such computational design, the design problem is rigorously formulated as a mathematical problem, which is then solved through optimization analysis. Compared to the traditional trial-and-error design approach, its strengths are evident, such as implementation autonomy without manual interference, excellent capability of solution search, and so on. Within this computational framework, previous studies have analytically calculated the function related to the normal sound absorption coefficient and used it as the design optimization objective to guide the design analysis. Cobo et al. [42] applied a simulated annealing (SA) algorithm to optimize the parameters of triple-layer MPPs for the maximum mean absorption coefficient within a prescribed frequency band. Qian et al. [43] proposed a multi-population genetic algorithm (GA) to optimize the design of an MPP with varying hole sizes, aiming to maximize average absorption within a specified frequency band. Lu et al. [44] used the particle swarm optimization (PSO) algorithm to enhance the mean absorption of four distinct types of MPPs. Wang and Bennett [45] applied the sequential quadratic programming (SQP) algorithm to minimize the sound power reduction factor of an MPP featuring multiple chambers. In addition to the analytical derivation of the design optimization objective, finite element (FE) simulation has also been adopted for its greater accuracy and applicability in general cases [46-48]. Although the methods for evaluating the design optimization objective differ, the design implementation processes remain the same.

It is worth emphasizing that the state-of-the-art studies discussed above have primarily focused on examining a single design optimization objective, namely sound absorption performance. However, the structural parameters of multi-layer MPPs also significantly impact the absorber cost, particularly in terms of materials. For multi-layer MPPs to be widely employed in buildings, their cost also needs to be minimized. Moreover, multi-layer MPPs with different layers exhibit varying acoustic performances and assembly costs. Therefore, the number of layers should be considered an independent design variable in the optimization problem. Previous studies, on the other hand, have typically focused on MPPs with a fixed number of layers. Given this context, multi-objective design optimization with an expanded design space for multi-layer MPPs is essential for continuous advancements in sound absorption capacity and material cost. Nevertheless, this approach presents a technical challenge from a mathematical optimization perspective. Specifically, the number of layers is a discrete variable, while the MPP parameters, such as the diameter of perforations, panel thickness, perforation ratio, and cavity depth, are continuous variables. Moreover, the total number of MPP parameters depends on the number of layers. This combination makes the design problem in this study a hybrid optimization problem with constraints on different types of design variables. To the best of the authors' knowledge, such a design problem holds practical significance but has not yet been thoroughly investigated or solved.

To address the aforementioned challenges and achieve the effective MPP design, this study aims to establish an integrated design framework. This framework is built upon the seamless integration of numerical modeling, experimental investigation, and novel multi-objective optimization algorithm to simultaneously maximize sound absorption capacity and minimize material cost. In-house codes are specifically developed for the multi-objective optimization algorithm, as well as for the integration and implementation of the framework. The entire manuscript is structured as follows: Section 1 presents the background of the study. Section 2 establishes the finite element (FE) model used to conduct the acoustic analysis of the MPP, followed by thorough experimental validation. Section 3 formulates the design optimization problem and introduces the proposed multi-objective optimization algorithm to facilitate the multi-layer MPP design. Section 4 draws conclusions and offers future prospects.

## 2. Numerical modeling with experimental validation

### 2.1. Acoustic analysis of MPPs

As illustrated in Fig. 2, the general configuration of the MPP is defined by its geometric parameters, including the diameter of the perforation (i.e., $d$), the panel thickness (i.e., $t$), and the cavity depth (i.e., $D$). For a cylindrical MPP with the radius of $A$ and the $N$ circle holes uniformly distributed over the panel, the perforation ratio $\sigma$, defined as the ratio of perforations' area to the total surface area of the panel, can be calculated as

$$\sigma = \frac{Nd^2}{4A^2} \tag{1}$$

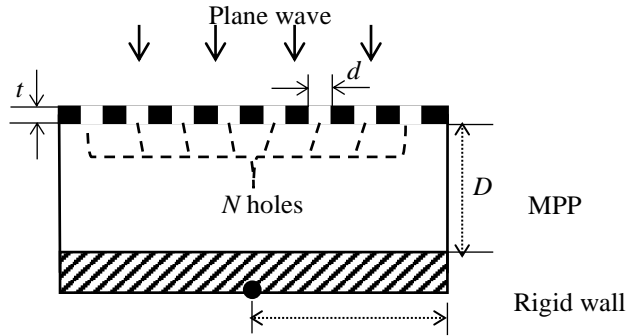

**Fig.2.** Illustration of the general MPP configuration.

After specifying the above geometric parameters, the numerical model can be established through the finite element (FE) method, and the acoustic analysis can then be performed. As illustrated in Fig. 3, the model typically consists of three subdomains, including perfectly matched layer (PML), waveguide, and MPP subdomains, intending to elucidate the sound wave propagation characteristics. Specifically, PML in this model is imposed to absorb outgoing waves from the wave subdomain. The waveguide and MPP subdomains are backed by the hard/rigid walls. The sound propagation in the frequency domain is characterized by the Helmholtz equation as

$$\nabla \cdot \left(-\frac{1}{\rho}\nabla p\right) - \frac{p\omega^2}{\rho c^2} = 0 \tag{2}$$

where $\rho$ denotes the air density, $p$ denotes the sound pressure, $\omega$ denotes the angular frequency of sound wave, and $c$ denotes the sound speed in the air. By solving Eq. (2), the total acoustic pressure can be expressed as

$$p_t = p_i + p_r \tag{3}$$

where $p_i$ and $p_r$ denote the incident and reflected acoustic pressure, respectively. $p_i$ can be derived as

$$p_i = p_0 exp(-j\mathbf{k} \cdot \mathbf{x}) \tag{4}$$

where $p_0$ denotes the wave amplitude, $\mathbf{k}$ denotes the wave vector defined by the wave number, and $\mathbf{x}$ represents the location on the interior boundary. When the sound wave passes through an interior boundary, the acoustic pressure loss can be described as

$$\Delta p_t = Z_t \cdot v_n \tag{5}$$

where $Z_t$ represents the transfer impedance of an interior boundary given a medium or material, and $v_n$ represents the normal acoustic velocity. Here, transfer impedance measures the ability of a boundary or structure to transmit sound energy from one side to the other. This is a key parameter for describing the sound wave propagation behavior through a medium or material and should therefore be considered in acoustic analysis.

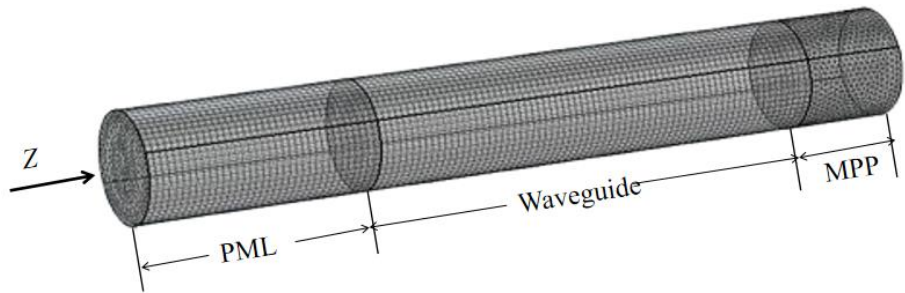

**Fig.3.** Description of FEM model domains.

The transfer impedance of the MPP is influenced by several factors, including the transfer impedance of holes, end correction, hole-hole interaction, nonlinear effects, and resonance of air cavity. Because the panel is very thin, its energy losses due to heat conduction are negligible. As shown in Fig.4(a), we assume the sound wave is injected along the z-coordinate axis perpendicular to the panel. The acoustic pressure and the sound velocity along the z-axis are

$$p(z) = p\exp(-jk_c z), v(z) = v\exp(-jk_c z) \qquad (6a, b)$$

where $k_c$ denotes the complex wave number. Substituting Eqs. (6a, b) into Eq. (5) leads to

$$Z_n = \frac{p(z+t)-p(z)}{v_n(z+t/2)} = -2jZ_c \sin\frac{k_c t}{2} \qquad (7)$$

where $Z_n$ represents the transfer impedance of a hole, and $Z_c$ represents the characteristic impedance, which is the product of the density of the medium and the speed of sound within it. Since thermal effects are negligible for thin panel, the normalized transfer impedance of a hole can be expressed based on Eq. (7) as follows [49]:

$$z_n' = \frac{Z_n}{\rho c} = -\frac{j\omega}{c}\frac{t}{\gamma_v} \qquad (8)$$

where $\gamma_v$ denotes the mean value of the scalar viscous function at the cross section. The scalar viscous function typically represents viscosity at a given point in the fluid. Consequently, the normalized transfer impedance of all the holes is described as

$$z_p = \frac{z_n'}{\sigma C_D} \qquad (9)$$

where $C_D$ represents the ratio of the actual flow rate of a fluid through a hole to the theoretical flow rate under ideal conditions.

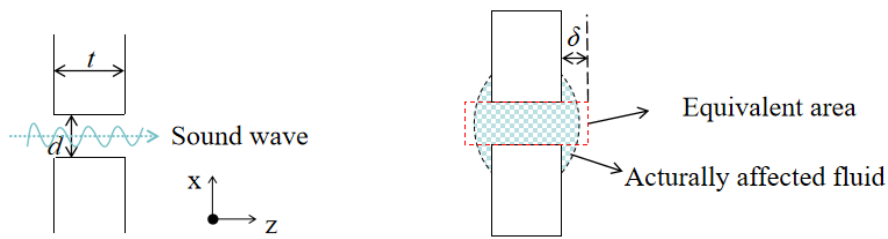

(a) (b)

**Fig.4.** Propagation of sound waves: (a) direction of sound waves; (b) rationale behind end correction.

It should be noted that Eq. (9) only considers the fluid within the holes. However, the incident wave affects more fluid around the hole. The equivalent area is longer than the panel thickness by $\delta$ on each side, as shown in Fig.4(b). Therefore, Eq. (9) should be modified to account for end corrections for the equivalent area of the fluid. The hole-hole interaction can be evaluated by a factor $f_{int}$, the so-called the *Fok function*, which is expressed by

$$f_{int} = \sum_{n=0}^{8} a_n (\sqrt{\sigma})^n \tag{10}$$

where $a_n$ denotes the multinomial coefficients. Hence, the end correction and hole-hole interaction can be further expressed as

$$z_{end} = -\text{Re}(\frac{j\omega}{c\sigma C_D} \frac{2\delta}{\gamma_v} f_{int}) - j\text{Im}(\frac{j\omega}{c\sigma C_D} \frac{2\delta}{\gamma_v} f_{int}) \tag{11}$$

The flow separation and vortex shedding induced by medium and high sound pressure levels increase the nonlinear acoustic resistance, which is as follows [50]:

$$z_{nl} = \frac{1-\sigma^2}{\sigma^2 C_D^2} \frac{1}{2c} |v_n| \tag{12}$$

Moreover, the normalized transfer impedance of the back air cavity can be calculated as [51,52]

$$z_{air} = -j\cot(k_c \cdot D) \tag{13}$$

Combining Eqs. (8)-(13), the normalized transfer impedance of the MPP can be derived as

$$z_t = -\text{Re}(\frac{j\omega}{c\sigma C_D}l \frac{t+2\delta}{\gamma_v} f_{int}) - j\text{Im}(\frac{j\omega}{c\sigma C_D}l \frac{t+2\delta}{\gamma_v} f_{int}) + \frac{1-\sigma^2}{\sigma^2 (C_D{}^{nl})^2} \frac{1}{2c} |v_n| - j\cot(k_c \cdot D) \tag{14}$$

where $z_t$ denotes the normalized transfer impedance of the MPP. On this basis, the sound absorption coefficient (SAC), which represents sound absorption ability of the MPP, is derived as [51-53]

$$\alpha = 1 - \left|\frac{z_t-1}{z_t+1}\right|^2 \tag{15}$$

where $\alpha$ denotes the SAC. As can be seen from the above formulation, the SAC is essentially a function of geometric parameters depicted in Fig. 2. As an important measure of acoustic performance, it will be used as the design optimization objective to guide the MPP design, as will be introduced in Section 3.1.

To implement the fundamental principle outlined above, we utilize *COMSOL Multiphysics* to build the finite element (FE) model and conduct the acoustic analysis. The perfectly matched layer (PML) and waveguide subdomains are discretized by swept meshing, while tetrahedral elements are employed for the MPP subdomain. The aluminum is specifically chosen for the MPP, including both the panel and cavity, according to its wide use [54,55]. The sizes of all elements are set below one-sixth of the minimum wavelength of the incident sound wave to ensure the desired simulation accuracy. The entire computational domain in FE model consists of over 44,000 degrees of freedom (DOFs). A single FE simulation run takes approximately 2 minutes to complete on a computer with an i5-8400 CPU, 4 GB of RAM memory, and the Windows 10 Home operating system.

## 2.2. Experimental correlation analysis

To validate the accuracy of the FE model established in Section 2.1, an experimental investigation is conducted. Specifically, two MPP specimens with varying perforation ratios and cavity depths (as shown in Fig. 5) are fabricated, in which the MPP panels are perforated using the electrical discharge machining (EDM). The panel thickness and perforation diameter of both specimens are 0.35 mm and 0.25 mm respectively. The short MPP, with a cavity of 6 cm, has a perforation ratio of 0.75%, while the tall MPP, with a cavity of 9 cm, has a perforation ratio of 0.9%. It is noted that the specimens with the abovementioned geometric parameters were chosen based on the practical implementation experience and the specific installation environment. The impedance tube test for these two specimens is carried out.

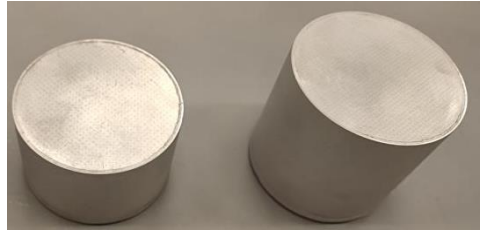

**Fig.5.** Two MPP testing specimens.

In this test, a two-microphone impedance tube (*Brüel & Kjær Impedance Tube Kit Type 4206*) following the *ISO 10534-2* and *ASTM E1050-12* standards is utilized to measure sound absorption coefficient (SAC) of specimens. Since few indoor noise sources exhibit significant levels above 1600 Hz that pose a substantial concern in residential areas [56,57], we focus on the sound absorption performance of MPPs below 1600 Hz. Therefore, we use a 100 mm diameter tube of Type 4206, which measures frequencies from 50 Hz to 1600 Hz, and design the external diameter of specimens accordingly for installation (see Fig. 5). Before testing, microphone calibration is conducted to ensure the reliability and accuracy of the results. After gently inserting the sample into the tube, a final inspection verifies a perfect fit without visible indentations. A loudspeaker at one end of the tube generates broadband stationary random sound waves, and two microphones measure the resulting sound pressure. The SAC of specimens is derived based on *ISO 10534-2* and *ASTM E1050-12* methods. Each specimen is tested three times to minimize the adverse effect of uncertainties, and the recorded values are averaged. The test equipment layout is shown in Fig. 6.

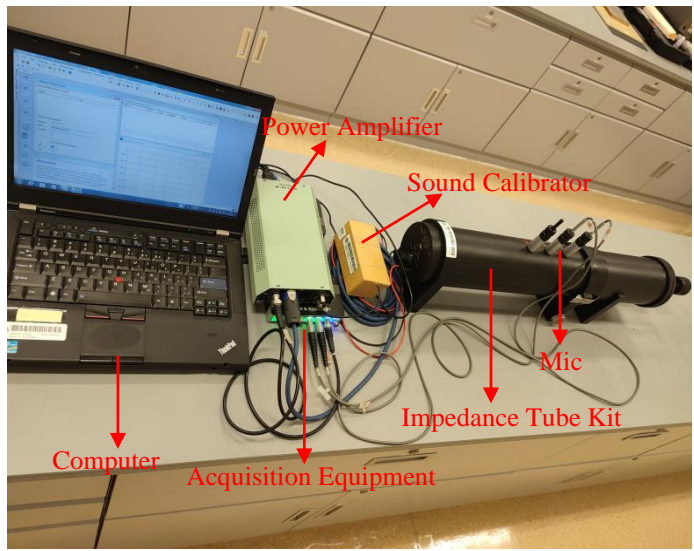 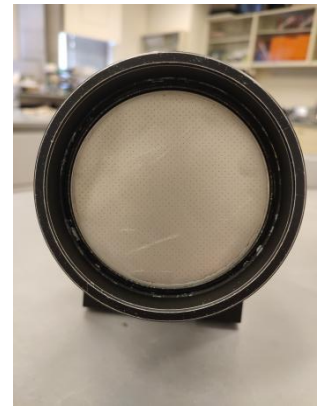

(a) (b)

**Fig. 6.** The impedance tube test: (a) instrumentation setup; (b) the mounted specimen inside the tube.

The performance of the MPPs for these two specimens has been further analyzed. The results, shown in Fig. 7, compare the experimental data (blue line) with the numerical results (red line). As shown, the specimen in Fig. 7b outperforms that in Fig. 7a within the frequency range below approximately 700 Hz. Conversely, as the target frequency increases, the specimen in Fig. 7a exhibits a relatively slower decrease in performance beyond 700 Hz. The frequency ranges corresponding to the absorption peaks show slight shifts between these two specimens. However, more pronounced frequency shifts for absorption peaks are expected among the design candidates under combinatory design in high-dimensional space. The notable discrepancies in performance, as shown in Fig. 7, emphasize the necessity of optimizing MPP design across the entire target frequency range.

Interestingly, sound absorption valleys are observed in the experimental results. This phenomenon is primarily due to panel vibrations that compromise absorption performance at the structural resonant frequency [58,59]. This complex effect is not evident in the numerical results, as the panel is considered ideally rigid in the finite element (FE) model to reduce computational complexity and thus facilitate FE-based inverse optimization analysis. However, the numerical results overall align with the experimental results, except for the presence of the sound absorption valley, indicating the capability of the FE model to capture the underlying acoustic characteristics of actual MPPs. Furthermore, it is worth noting that the presence of the sound absorption valley may play an insignificant role in the subsequent MPP design optimization, given its possibly consistent occurrence across all cases. The comparison of results verifies the FE model accuracy, laying the foundation for subsequent MPP design optimization, which leverages the high-fidelity FE model as the cornerstone.

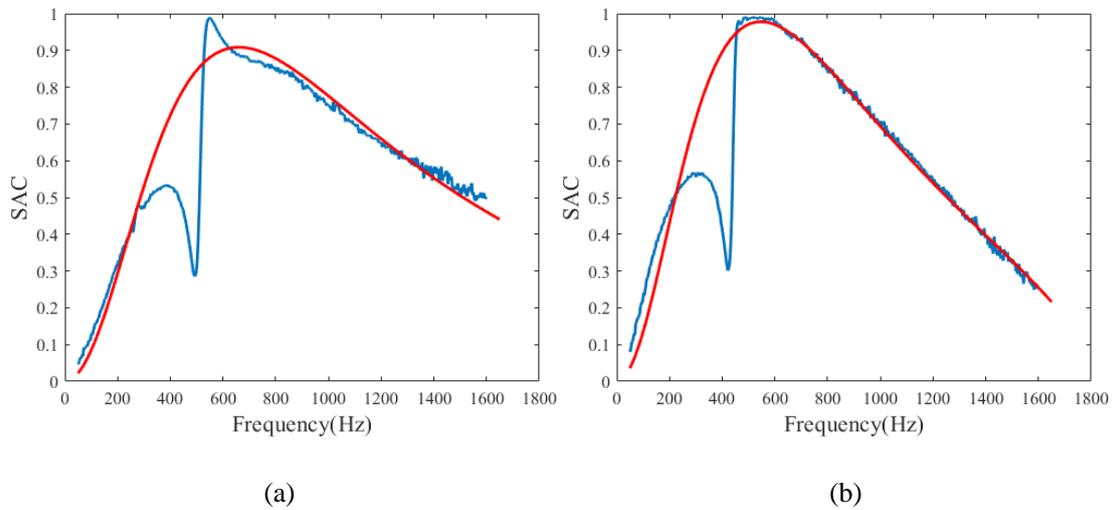

(a)  (b)

**Fig.7.** Comparison of experiment and numerical results for MPPs with a thickness of 0.35 mm and perforation diameter of 0.25 mm: (a) $\sigma$=0.75%, $D$=6 cm; (b) $\sigma$=0.9%, $D$=9 cm.

## 3. Design optimization implementation

### 3.1. Problem formulation

As introduced in Section 2.1, the sound absorption coefficient (SAC) of the MPP depends on its geometric parameters of ($d, t, \sigma, D$). For an MPP with $i$ layers, as shown in Fig. 8, its geometric parameters can be generally represented as a vector of ($d_1, t_1, \sigma_1, D_1, d_2, t_2, \sigma_2, D_2,..., d_i, t_i, \sigma_i, D_i$). Increasing the number of layers, i.e., $i$, will certainly expand the design optimization space, therefore compounding the complexity of optimization analysis. As a special case, the multi-layer MPP converts into a single-layer MPP when $i$ is set to 1.

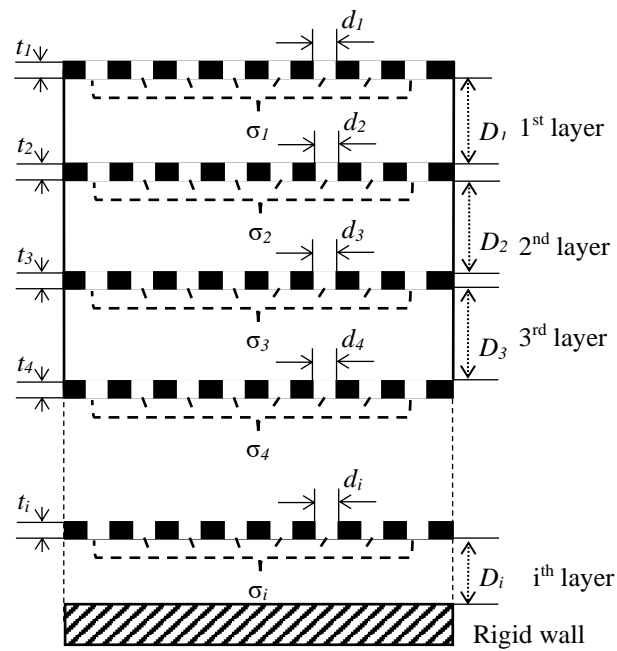

**Fig.8.** Illustration of the multi-layer MPP configuration.

The effectiveness of MPP is primarily measured by its sound absorption performance. As mentioned in the Introduction, low-frequency noise is the major component of indoor noise that disrupts residents. Therefore, this research focuses on enhancing absorption performance below 1600 Hz, aligning with the impedance tube testing capacity described in Section 2.2. Without bias towards specific frequency components, we formulated a metric to evaluate the overall absorption performance of the MPP by integrating the SAC curve over the target frequency range of 50-1600 Hz. This comprehensive metric considers both the frequency attenuation band and amplitude reduction, which is given as [46,47]

$$S = \int_{50}^{1600} \alpha(f) \approx \sum_{50}^{1600} \alpha(f) \tag{16}$$

where $\alpha(f)$ denotes the SAC value at a certain frequency $f$. SAC curve here can be evaluated through finite element (FE) simulation given the MPP configuration. This performance indicator can be written as an implicit function of geometry parameters:

$$S = \vartheta(d1, t1, \sigma1, D1, d2, t2, \sigma2, D2, \ldots, di, ti, \sigma i, Di) \tag{17}$$

where $\vartheta$ indicates the operation involving the FE simulation and SAC curve integration.

For practical deployment in building environments, the fabrication cost of the MPP from an economic perspective should also be considered. In this research, the fabrication cost of the MPP is defined as the sum of material and processing costs. Recall that a typical MPP consists of layers of panels, a cavity, and a rigid wall, as illustrated in Fig. 8. Since MPPs are usually placed in front of walls or beneath ceilings [1], there is no need to specially fabricate the rigid wall. For MPPs applied over a wide area, the fabrication cost of the cavity is often much less than that of the panels. Therefore, the fabrication cost of the MPP is mainly covered by the cost of the panels, which can be simply estimated as

$$C_f = C_m + C_p \tag{18}$$

where $C_f$, $C_m$, and $C_p$ represent the fabrication, material, and processing costs of the panels included in the MPP, respectively. The total cost of materials used for fabricating panels can be calculated as

$$C_m = \sum_i A \cdot t_i \cdot \rho \cdot P_m \tag{19}$$

where $A$ denotes the panel surface area, which is typically the same for all panels. $t_i$ is the thickness of the $i$-th panel, $\rho$ is the density, and $P_m$ is the unit price of the panel material.

Based on empirical knowledge, the processing cost of panels is approximately proportional to the number of holes in them, as expressed by the following equation:

$$C_p = \sum_i N_i \cdot P_h \tag{20}$$

where $N_i$ is the hole number of the $i$-th panel, which can be derived based on the respective perforation ratio defined in Eq. (1). $P_h$ denotes the unit price of perforating a hole by electrical discharge machining (EDM) method.

With the full definition of costs in relation to materials and processing (Eqs. (19) and (20)), the total fabrication cost of the MPP outlined in Eq. (18) can be evaluated. Given that aluminum is primarily used for the panels, its material properties and unit price are well-documented. Specifically, the density of aluminum, i.e., $\rho$ in Eq. (19) is set at $2.7 \times 10^3$ kg/m³, and the unit price, i.e., $P_m$ is approximately 20 HKD/kg. The surface area of the panel, i.e., $A$ in Eq. (19) is consistently set at 1 m². Moreover, the price for processing a single hole, i.e., $P_h$ in Eq. (20) is approximately 0.04 HKD based on the consultation from the manufacturing company. Although the unit prices used may not be entirely precise, they do not impact the optimization results as they remain constant. We select as reasonable price values as possible to ensure the actual costs are easy to interpret in the succeeding result discussion. By substituting all constants into the Eqs. (19) and (20), the Eq. (18) can be rewritten as

$$C_f = \sum_i [5.4 \times 10^4 \, t_i + \frac{0.16 \sigma_i}{\pi d_i^2}] \tag{21}$$

The two performance indicators presented in Eqs. (17) and (21) are inherently related to the parameters ($d$, $t$, $\sigma$, $D$). Therefore, these parameters can be considered as design variables in the subsequent design optimization. It is well known that these parameters should fall within a certain range to well serve the purpose of practical implementation. For example, the panel of the MPP is ultra-thin, usually no more than 1 mm in thickness [1,22], with holes whose diameters are also no more than 1 mm [60]. Typically, the perforation ratio of the MPPs is under 2% [1,23,39]. The total cavity depth depends on the specific installation environment. In this study, we assume that the total thickness of the MPP should not exceed 9 cm, with the total cavity depth playing a dominant role. Based on the above set of constraints, the multi-objective optimization problem, specifically a bi-objective optimization problem involving the design objectives shown in Eqs. (17) and (21), is formulated as follows:

$$\begin{cases} \text{Maximize } S(x) \\ \text{Minimize } C_f(x) \\ x = [d_1, t_1, \sigma_1, D_1, d_2, t_2, \sigma_2, D_2, \dots, d_i, t_i, \sigma_i, D_i, i] \end{cases}$$

Subject to

$$0.0001 \leq d_i \leq 0.001 \text{ m}, i \in N^*$$

$$0.0001 \leq t_i \leq 0.001 \text{ m}, i \in N^*$$

$$0.002 \leq \sigma_i \leq 0.02, i \in N^* \quad (22)$$

$$D_i > 0, i \in N^*$$

$$\sum_i D_i + t_i \leq 0.09 \text{ m}, i \in N^*$$

where $N^*$ is the maximum number of layers in the MPP. In this research, we limit the maximum number of layers to three to ensure the system development remains cost effective in terms of fabrication and assembly. To facilitate the optimization, the negative $S(x)$ will be used to convert the problem into a minimization problem.

## 3.2. Multi-objective particle swarm optimization algorithm and integrated design framework

According to the problem described in Eq. (22), the number of layers in the MPP is essentially a discrete variable denoted by *i*. This discrete variable dynamically influences the number of continuous design variables, as each layer contains four holes with radii treated as continuous values. Consequently, changes in the number of layers lead to corresponding changes in the number of design variables. The challenge lies in managing both the hybrid nature of variables (discrete and continuous) and dynamic adjustments to these variables during the optimization process. To address these issues, we develop an alternating optimization algorithm on the basis of the multi-objective particle swarm optimization (MOPSO), to solve the multi-objective optimization problem formulated in Eq. (22), while obtaining a set of trade-off solutions. Among the various optimizers, MOPSO is selected due to its rapid convergence and generation of a range of trade-off solutions in a single run. Additionally, MOPSO allows for the development of local search strategies that can further refine the search process. PSO algorithm is a stochastic technique first introduced by Eberhart and Kennedy [61]. It models the social behavior of animals such as insects, herds, birds, and fish. In this cooperative approach, each member of the swarm adjusts its search pattern based on both its own learning experiences and those of other members.

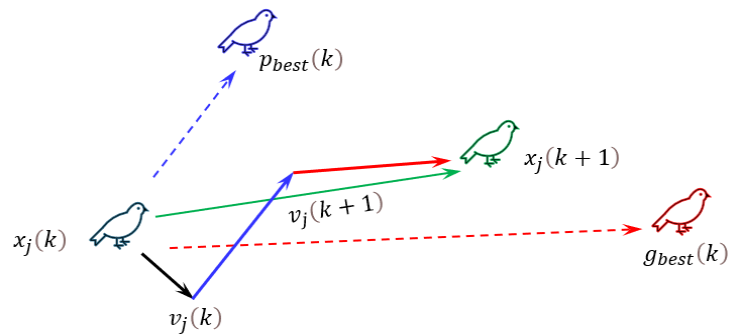

**Fig. 9**. Illustration of particle swarm-based search.

The search process of the PSO is illustrated in Fig. 9. In this process, each particle moves towards its personal best position stored in its memory and the global best position among all particles. Mathematically, the new velocity and position of each particle can be expressed as Eqs. (22) and (23):

$$v_i(t+1) = wv_i(t) + c_1r_1\big(p_{best}(t) - x_i(t)\big) + c_2r_2(g_{best}(t) - x_i(t)) \tag{23}$$

$$x_i(t+1) = x_i(t+1) + v_i(t+1) \tag{24}$$

where $v$ is the velocity of the particle, $x$ is the position of particle or the solution of the optimization problem, $r_1$ and $r_2$ are random numbers, $t$ is the current iteration, $p_{best}$ is the personal best, and $g_{best}$ is the global best. In the above equations, coefficient $w$ is the inertia weight, $c_1$ is the coefficient of the cognitive component which indicates that each particle learns from its experience, and $c_2$ is the coefficient of the social component, from which all particles learn. Although PSO is used in various applications [62,63], handling problems with hybrid and dynamic variables is challenging. Therefore, we adapt the conventional MOPSO by incorporating an alternative loop over the discrete variables. Additionally, the hypervolume indicator is used to compare the performance across different numbers of isolation layers. The hypervolume is a metric used in multi-objective optimization to measure the volume of the objective space dominated by a solution set, relative to a chosen reference point. It quantifies the performance of a solution by calculating the space it covers, with larger hypervolumes indicating better solutions in terms of convergence and diversity [64].

The algorithm operates by iterating over different discrete variable values (1, 2, 3). For each discrete variable value, a set of particles (solutions) is initialized, where each particle has 3, 6, or 9 continuous variables, depending on the chosen discrete variable. The MOPSO is then applied, allowing particles to move within the search space while updating their positions and velocities.

Each particle's objective function values are evaluated, and non-dominated solutions are stored in an archive. After each MOPSO run, the hypervolume of the current Pareto front is calculated. If the hypervolume improves, the corresponding discrete variable value and Pareto front are saved. This process is repeated multiple times to identify the optimal discrete variable value. Finally, the best discrete variable value and the corresponding Pareto front are returned.

This new algorithm integrated with further treatment mentioned above is referred to as alternating MOPSO in this study. It is developed using in-house code, and its implementation is demonstrated in the following pseudo code.

| **Algorithm:** Alternating MOPSO | |
|---|---|
| 1 | Initialize problem parameters: define cost function, set variable bounds |
| 2 | Set MOPSO parameters: maximum number of iterations, number of populations, coefficients for MOPSO |
| 3 | Alternating optimization loop: |
| 4 | **For** global _it =1 to *MaxGlobalIt*: |
| 5 |    **For** discrete _var = 1 to 3: |
| 6 |       -Set n _var based on discrete _var |
| 7 |       -Initialize population (random positions, velocities, constraints) |
| 8 |       -Evaluate particles and update personal bests |
| 9 |       -Determine non-dominated particles (add to repository) |
| 10 |       **For** it = 1 to *MaxIt:* |
| 11 |          **For** each particle: |
| 12 |             -Select leader from repository |
| 13 |             -Update velocity and position |
| 14 |             -Apply constraints |
| 15 |             -Recalculate costs |
| 16 |             -Apply mutation with probability |
| 17 |             -Update personal best if dominated |
| 18 |          **End For** |
| 19 |       -Add non-dominated particles to repository |
| 20 |       -Remove dominated particles and maintain repository size |
| 21 |    **End For** |
| 22 |       -Calculate hypervolume and track best results |
| 23 |    **End For** |
| 24 | **End For** |
| 25 | Output best results (discrete variable, hypervolume, Pareto front) |

With the proposed optimization algorithm and well-validated finite element (FE) model, the integrated design framework can be established to carry out the inverse optimization analysis, as depicted in Fig. 10. The solid line represents the outer layer, where the algorithm iterates over different discrete variable values corresponding to the number of layers. It is worth mentioning that the optimization algorithm is developed in *MATLAB*, while the FE simulation is conducted using *COMSOL Multiphysics*. Hence, performing autonomous optimization analysis necessitates the seamless integration of these development environments. Specifically, we developed a script to build the interface between these two environments. The interface allows the FE simulation to be activated and carried out backstage without manual graphic user interface (GUI) operation, and the analysis results to be exported to the algorithm for design objective assessment and subsequent sampling guidance. Through the automatic and iterative analysis during the design optimization process, the final non-dominant solutions can be identified under multi-objective framework.

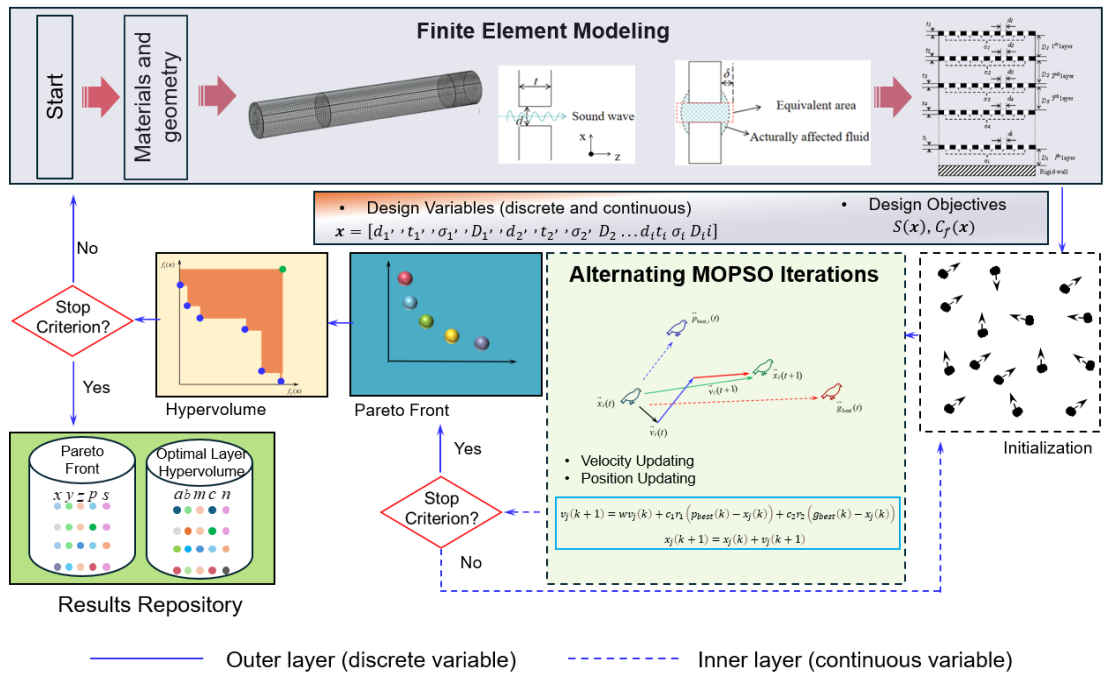

**Fig.10.** Integrated design framework.

### 3.3. Results and discussion

The inverse optimization analysis using the proposed MOPSO algorithm is conducted on a personal computer equipped with an i5-8400 CPU, 4 GB of RAM memory, and the Windows 10 Home operating system. As mentioned, a single finite element (FE) simulation run takes approximately 2 minutes. The operational parameters, as outlined in the pseudo code in Section 3.2, are set to 4 for maximum number of iterations, 20 for number of populations, 2 for coefficients for MOPSO. This setup is expected to ensure satisfactory optimization convergence with a tractable computational cost of approximately 10 hours. Therefore, no surrogate model is needed to replace the FE model, in order to ensure the high prediction accuracy that has already been validated in Section 2.2.

The Pareto solutions from the analysis are listed in Table 1, and their corresponding Pareto fronts are illustrated in Fig. 11. The performance measure, represented on the vertical axis, employs the negative $S(x)$ to adhere to the minimization problem. Totally, the 16 solutions exist in the Pareto solution set, including the optimized MPPs with different layers. Although the MPPs with single layer are generally inferior in terms of the performance (denoted by $S$), they are of less fabrication costs (denoted by $C_f$). When the number of layers in the MPP increases, the increasing complexity of manufacturing will notably elevate the cost. Along with the cost increase, the performance is significantly improved. As the number of layers in the MPP increases, the manufacturing complexity rises, leading to a notable increase in cost. However, this cost increase is accompanied by a significant improvement in performance. A 3-layer MPP solution (case 16) can boost performance by nearly 4-5 times compared to a 1-layer MPP (case 1), though it incurs 40 times the cost. The current solution set comprises nine 1-layer MPP solutions, six 2-layer MPP solutions, and one 3-layer MPP solution. This distribution implies that further increasing the layers of MPP may not be very effective when considering the balance between performance and cost. Given the stochastic nature of the optimization, additional runs may be necessary to elucidate the fundamental interrelation of optimized solutions in this design problem.

This solution set, derived from a multi-objective optimization framework, demonstrates increased flexibility in practical operations compared to a single solution from a single-objective optimization framework. With multiple choices available, engineers can integrate their preferences and judgment to select the best solution for a specific application context. For example, if performance is the top priority, increasing the number of layers at the expense of higher costs is worthwhile. Additionally, more objective methods, such as post-filtering and recommendation systems, can aid in achieving wise decision-making [65,66].

Table 1. Optimized geometric parameters in Pareto solutions ($d$ and $t$ are in millimeters, $\sigma$ is in percentage, $D$ is in centimeter, and $C_f$ is in HKD)

| Case | $d_1$ | $t_1$ | $\sigma_1$ | $D_1$ | $d_2$ | $t_2$ | $\sigma_2$ | $D_2$ | $d_3$ | $t_3$ | $\sigma_3$ | $D_3$ | $C_f$ | $-S$ |
|---|---|---|---|---|---|---|---|---|---|---|---|---|---|---|
| 1 | 0.91 | 0.44 | 0.20 | 6.12 | - | - | - | - | - | - | - | - | 148.22 | -13.80 |
| 2 | 0.70 | 0.33 | 0.20 | 5.26 | - | - | - | - | - | - | - | - | 224.86 | -17.83 |
| 3 | 1.00 | 0.59 | 0.20 | 4.11 | - | - | - | - | - | - | - | - | 133.97 | -12.22 |
| 4 | 0.83 | 0.44 | 0.41 | 2.20 | - | - | - | - | - | - | - | - | 328.76 | -22.96 |
| 5 | 0.71 | 0.16 | 0.44 | 4.34 | - | - | - | - | - | - | - | - | 462.13 | -28.05 |
| 6 | 0.55 | 0.67 | 0.46 | 2.18 | - | - | - | - | - | - | - | - | 820.90 | -28.18 |
| 7 | 0.42 | 0.13 | 0.49 | 5.83 | - | - | - | - | - | - | - | - | 1394.5 | -40.24 |
| 8 | 0.55 | 0.55 | 0.56 | 4.95 | - | - | - | - | - | - | - | - | 965.41 | -30.78 |
| 9 | 0.75 | 0.38 | 0.84 | 4.64 | - | - | - | - | - | - | - | - | 772.19 | -28.14 |
| 10 | 0.67 | 0.38 | 1.07 | 1.74 | 0.55 | 0.27 | 0.44 | 3.17 | - | - | - | - | 1976.7 | -48.08 |
| 11 | 0.36 | 0.48 | 1.09 | 2.07 | 0.75 | 0.10 | 0.24 | 4.38 | - | - | - | - | 4502.5 | -53.20 |
| 12 | 0.61 | 0.70 | 1.10 | 3.80 | 0.76 | 0.31 | 0.20 | 2.05 | - | - | - | - | 1745.7 | -43.70 |
| 13 | 0.77 | 0.63 | 1.19 | 6.32 | 0.81 | 0.92 | 0.38 | 1.95 | - | - | - | - | 1396.3 | -40.35 |
| 14 | 1.00 | 0.68 | 1.37 | 3.86 | 0.58 | 0.24 | 0.20 | 1.20 | - | - | - | - | 1044.6 | -33.34 |
| 15 | 0.66 | 0.71 | 0.73 | 5.44 | 0.96 | 0.60 | 0.20 | 3.56 | - | - | - | - | 1046.6 | -35.96 |
| 16 | 0.66 | 0.11 | 0.90 | 3.75 | 0.39 | 0.79 | 0.96 | 2.64 | 0.47 | 1 | 0.78 | 2.61 | 6256.9 | -61.14 |

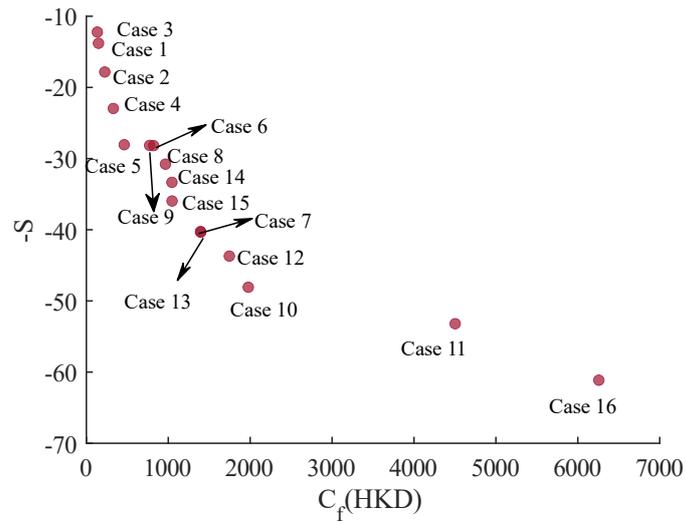

Fig.11. Pareto fronts corresponding to optimized solutions.

## 4. Conclusion

The low-frequency components of indoor noise in buildings significantly degrade residential quality. Although microperforated panels (MPPs) have been proven effective in mitigating low-frequency noise, their size, fabrication process, and associated costs need to be considered during development to ensure practical viability. In this paper, we focus on designing MPPs to optimize both sound absorption performance and fabrication cost through an integrated design framework. A finite element (FE) model of the MPP is established and validated using impedance tube tests to accurately predict the sound absorption coefficient (SAC) of MPPs under different operating conditions. A novel multi-objective particle swarm optimization (MOPSO) algorithm is then developed to conduct inverse analysis-based design, utilizing the above well-established MPP FE model. This algorithm can effectively address the technical challenges arising from the nature of the design problem, which involves both discrete and continuous design variables with correlations. An in-house code is developed to facilitate the entire design implementation, yielding a set of non-dominant solutions. The results offer the useful guidance for decision making, demonstrating the feasibility of the proposed methodology.

Future endeavors will primarily focus on exploring the use of flexible panels in MPP design by incorporating additional design variables, such as panel area, which interrelates with the flexible panel. The consideration of the flexible panel in the MPP also introduces the complex structural-acoustic coupling effect, which will increase the computational burden. Therefore, it is essential to investigate computational expedition strategies. More specific design objectives regarding performance and cost can be formulated. For example, instead of examining the sound absorption capacity over the entire frequency range, we can set multiple objectives to assess the capacity at different sub-frequency ranges defined by the users. This approach ensures a tailored design purpose. Additional efforts will be made to further advance optimization algorithms to handle more complex and general design problems of MPPs and other sound absorption materials.

## Acknowledgement

The research is supported part by the start-up fund provided by The Hong Kong Polytechnical University (PolyU), and part by the research project fund from Research Institute for Sustainable Urban Development (RISUD) at PolyU.